\documentclass[prl,twocolumn,showpacs,twoside,preprintnumbers,amsmath,amssymb]{revtex4}

\usepackage{graphicx,color}
\usepackage{dcolumn}
\usepackage{bm}

\begin{document}

\preprint{CAS-KITPC/ITP-126}

\title{Microscopic Analysis of Order Parameters in Nuclear Quantum Phase Transitions}

\author{Z. P. Li$^{1,2}$}
\author{T. Nik\v{s}i\'{c}$^{2}$}
\author{D. Vretenar$^{2,3}$}
\author{J. Meng$^{1,3,4}$}

\affiliation{$^{1}$State Key Laboratory of Nuclear Physics and Technology, School of Physics,
Peking University, Beijing 100871, China}
\affiliation{$^{2}$Physics Department, Faculty of Science, University of Zagreb,
10000 Zagreb, Croatia}
\affiliation{$^{3}$ Kavli Institute for Theoretical Physics China, CAS, Beijing 100190, China}
\affiliation{$^{4}$School of Physics and Nuclear Energy Engineering,
Beihang University, Beijing 100191, China}

%

\date{\today}

\begin{abstract}
Microscopic signatures of nuclear ground-state shape phase transitions in Nd isotopes
are studied using excitation spectra and collective wave functions obtained by diagonalization
of a five-dimensional Hamiltonian for quadrupole vibrational and rotational degrees of freedom,
with parameters determined by constrained self-consistent relativistic mean-field calculations for
triaxial shapes. As a function of the physical control parameter -- the number of nucleons, energy
gaps between the ground state and the excited vibrational states with zero angular momentum,
isomer shifts, and monopole transition strengths, exhibit sharp discontinuities at neutron number
$N=90$, characteristic of a first-order quantum phase transition. 
\end{abstract}

\pacs{21.60.Jz, 21.60.Ev, 21.10.Re, 21.90.+f}

\maketitle

Phase transitions in equilibrium shapes of atomic nuclei correspond
to first- and second-order quantum phase transitions (QPT) between
competing ground-state phases induced by variation of a non-thermal
control parameter (number of nucleons) at zero
temperature. Theoretical studies have typically been based on
phenomenological geometric models of nuclear shapes and potentials,
or algebraic models of nuclear structure  
\cite{Iac.03,Rick.07,CJ.08}, but more recently several
attempts have been made towards a fully microscopic description of
shape QPT starting from nucleonic degrees of 
freedom \cite{Meng.05,SG.05,FBL.06,Ni.07,Li.09,RS.07,RE.08,RRS.08}. In particular,
in Refs.~\cite{Ni.07,Li.09} we have reported a microscopic study of
nuclear QPT in the region $Z = 60$, $62$, $64$ with $N \approx 90$,
based on constrained self-consistent relativistic mean-field calculations
of potential energy surfaces. While in Ref.~\cite{Ni.07} the generator
coordinate method (GCM) was used to perform configuration
mixing of angular-momentum and particle-number projected relativistic
wave functions restricted to axial symmetry, in \cite{Li.09}  collective excitation spectra
and transition probabilities have been calculated starting from a five-dimensional
Hamiltonian for quadrupole vibrational and rotational degrees of
freedom, with parameters determined by constrained mean-field calculations for
triaxial shapes, i.e. including both $\beta$ and $\gamma$ deformations.
The results reproduce available data, and show that there is an abrupt change of
structure at $N=90$ that can be approximately characterized by the
X(5) analytic solution at the critical point of the
first-order quantum phase transition between spherical and
axially deformed shapes.

A phase transition is characterized by a significant variation of one or more order
parameters as functions of the control parameter. Even though in systems composed
of a finite number of particles, i.e. in mesoscopic systems, phase transitions
are actually smoothed out, in many cases clear signatures of abrupt changes of
structure properties are observed. In their study of QPT transitions in mesoscopic
systems \cite{Iachello04}, Iachello and Zamfir have shown that the main features
of phase transitions, defined for an infinite number of particles, $N \to \infty$, persists
even for moderate $N \approx 10$. Their analysis has been followed by a number of
studies of shape phase transitional patterns in nuclei as functions of the number
of particles, e.g. number of bosons in the framework of interacting-boson-type
models. As emphasized in Ref.~\cite{Iachello04}, there are two approaches to
study QPT: (i) the method of Landau, based on potentials, and (ii) the direct
computation of order parameters. In the case of atomic nuclei, however, a
quantitative analysis of QPT must go beyond a simple study of potential energy surfaces.
This is because potentials or, more specifically, deformation parameters that characterize
potential energy surfaces, are not observables, and can only be related to observables by making
very specific model assumptions. The direct computation of 
observables related to order parameters has so far been based mostly on 
particular nuclear structure models, e.g. the interacting
boson model, in the framework of which such observables are defined 
as expectation values of suitably chosen operators. 
\begin{figure*}[htb]
\includegraphics[scale=1.1]{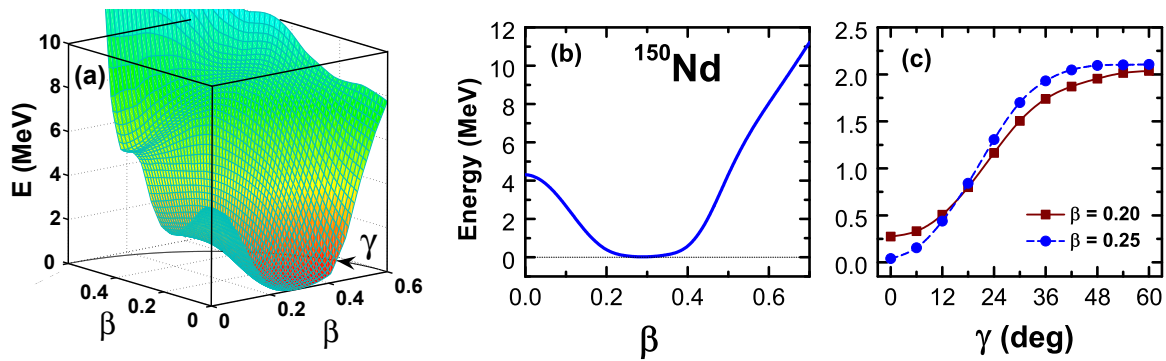}
\caption{\label{fig:PES3D-Nd} (Color online) (a) Self-consistent RMF+BCS triaxial
quadrupole binding energy map of $^{150}$Nd
in the $\beta - \gamma$ plane ($0\le \gamma\le 60^0$).
The contours join points on the surface with the same energy (in MeV).
(b) Projection of the binding energy map on the $\gamma = 0^0$ axis. (c) 
The dependence of the binding energy on the deformation parameter $\gamma$, 
for two values of the axial deformation $\beta =0.2$ and $0.25$. }
\end{figure*}

In this work we combine both approaches in 
a consistent microscopic framework, and present an illustrative 
example of calculation of observables that can be related to 
quantum order parameters as functions of nucleon number. An order 
parameter is a measure of the degree of order in a system. As a normalized 
quantity that is zero in one (symmetric) phase, and non-zero in the other, 
it characterizes the onset of order at the phase transition. When symmetry 
is broken, several variables can be introduced, related to order parameters, 
to describe the state of the system.
As in our previous studies \cite{Ni.07,Li.09}, the shape transition in
Nd isotopes with $N \approx 90$ will be considered. The analysis starts by performing
constrained self-consistent relativistic mean-field calculations for triaxial shapes, i.e. including
both $\beta$ and $\gamma$ deformations. The resulting self-consistent solutions, i.e. single-particle
wave functions, occupation probabilities, and quasiparticle energies that correspond to each point on
the binding energy surface, are used to calculate the parameters that determine the collective
Hamiltonian: three mass parameters, three moments of inertia, and the zero-point energy
corrections, as functions of the deformations $\beta$ and  $\gamma$ \cite{Ni.09}.
The diagonalization of the Hamiltonian yields the excitation energies and collective wave 
functions, that are used to calculate observables.

In Fig.~\ref{fig:PES3D-Nd} we plot the self-consistent triaxial
quadrupole binding energy map of $^{150}$Nd
in the $\beta - \gamma$ plane ($0\le \gamma\le 60^0$). The calculations
are performed by imposing constraints on expectation
values of the quadrupole moments $\langle \hat{Q}_{2 0}  \rangle$ and
$\langle \hat{Q}_{2 2}  \rangle$, the relativistic functional PC-F1 (point-coupling
Lagrangian)~\cite{BMM.02} is used in the particle-hole channel, and a
density-independent $\delta$-force is the effective interaction in
the particle-particle channel, with pairing correlations treated in the BCS
approximation. As shown in \cite{Li.09}, the binding energy
maps of Nd isotopes show a gradual transition from lighter spherical 
nuclei towards the strongly prolate deformed $^{152}$Nd. Of particular
interest are nuclei around $^{150}$Nd, for which experimental evidence for
shape phase transitional behavior has been reported \cite{Rei.02}. $^{150}$Nd
is considered to be a good example of empirical realization of the X(5) model for
the critical point of first-order phase transition between spherical and
axially deformed shapes \cite{FI.01}.

The microscopic binding energy surface of $^{150}$Nd displays
a flat prolate minimum that extends in the interval $0.2 \leq \beta \leq 0.4$
of the axial deformation parameter (Fig. 1b), and a parabolic dependence on
$\gamma$ for $\gamma \leq 30^\circ$ in the region of the flat prolate
minimum (Fig. 1c). The flat bottom of the potential has been considered a
signature of possible phase transition. There
are, however, problems when one considers deformations
as possible order parameters of phase transitions \cite{RE.08}. First,
deformation parameters are not observables, and can only be linked
to observables, e.g. transition rates, excitation energies,
within the framework of a specific model. Second, in calculations that include 
both $\beta$ and $\gamma$ degrees of freedom, a phase transition cannot 
be characterized by the behavior of just one deformation parameter, e.g. $\beta$. 
Results obtained with the five-dimensional collective
Hamiltonian show that many properties of the excitation spectra are
affected by $\beta - \gamma$ coupling. Band-head excitation energies, energy
spacings within the bands, and transition strengths depend on the $\gamma$
stiffness of the potential \cite{caprio}.

\begin{figure}[htb]
\includegraphics[scale=0.325]{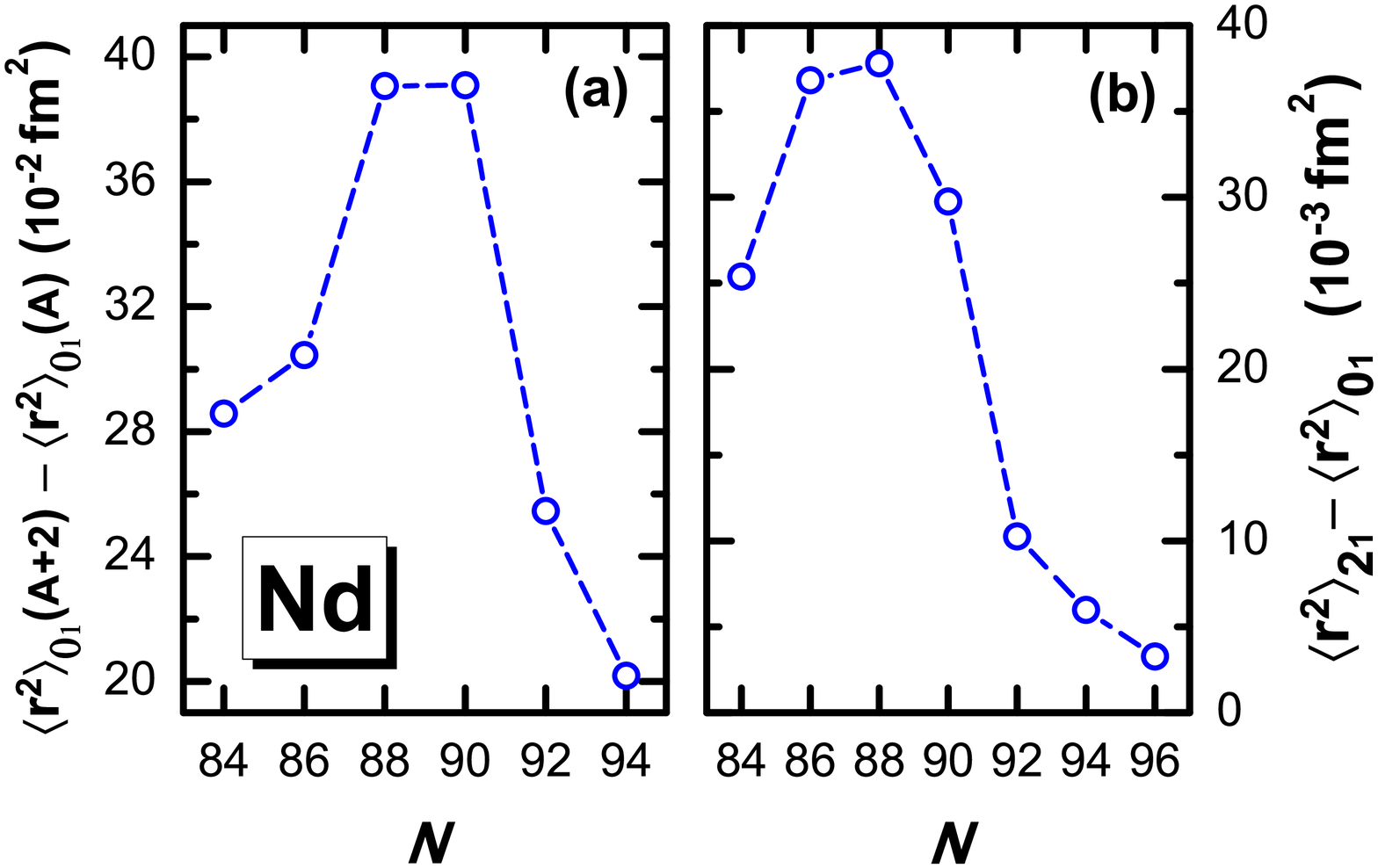}
\caption{\label{fig:Del-Rc-Nd} (Color online) Calculated differences 
between squares of ground-state charge radii:
$\langle r_c^2\rangle_{0_1^+}(A+2)-\langle r_c^2\rangle_{0_1^+}(A)$ (left panel),
and isomer shifts $\langle r_c^2\rangle_{2_1^+}-\langle r_c^2\rangle_{0_1^+}$ (right panel), as
functions of neutron number in Nd isotopes.}
\end{figure}

Here we analyze obervables that are
directly computed using collective wave functions obtained
from a microscopic five-dimensional collective Hamiltonian.
The important question is how much are the discontinuities at a phase transitional
point smoothed out in finite nuclei and, second, how precisely can a point of phase
transition be associated with a particular isotope, considering that the control parameter,
i.e. nucleon number, is not continuous but takes only discrete integer values. 
Fig.~\ref{fig:Del-Rc-Nd} displays the differences
between squares of ground-state charge radii  
$\langle r_c^2\rangle_{0_1^+}(A+2)-\langle r_c^2\rangle_{0_1^+}(A)$,
and the isomer shift $\langle r_c^2\rangle_{2_1^+}-\langle r_c^2\rangle_{0_1^+}$
between the first $2^+$ state and the ground state, as functions of the neutron number. 
The former displays a peak at $N=88,90$, whereas
a pronounced discontinuity is predicted for the latter between $N=88$ and $N=92$.

\begin{figure}[htb]
\includegraphics[scale=0.325]{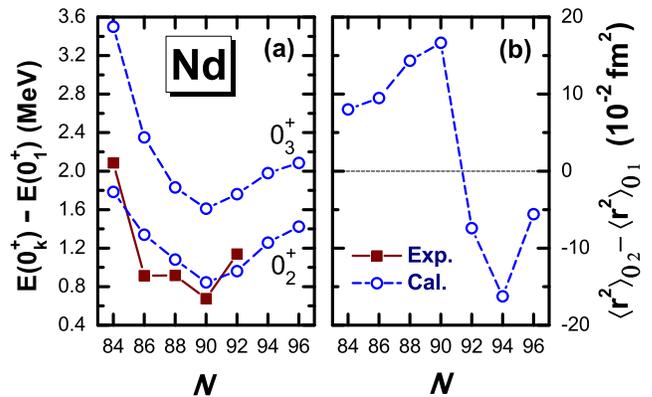}
\caption{\label{fig:E023-Rc-Nd} (Color online)
Evolution of the first and second excited $0^+$ states (left panel), and the isomer
shifts $\langle r_c^2\rangle_{0_2^+}-\langle r_c^2\rangle_{0_1^+}$ (right panel)
with neutron number in Nd isotopes. Microscopic values calculated with
the PC-F1 energy density functional are compared to data \cite{NNDC}.}
\end{figure}

Signatures of ground-state phase transitions in quantum systems characterize the
evolution of both excitation spectra and order parameters. In Refs.~\cite{Rowe.04,Dusuel.05,Arias.07}
the scaling properties of the energy gap between the ground state and the first excited vibrational
states with zero angular momentum was studied for a system of $N_B$ interacting bosons. At the
critical point of the phase transition the gap is strongly reduced in finite systems, and goes to zero as
$N \to \infty$. In the left panel of Fig.~\ref{fig:E023-Rc-Nd} we plot the isotopic dependence
of the first and second excited $0^+$ states in Nd nuclei and, in the right panel,
the isomer shift $\langle r_c^2\rangle_{0_2^+}-\langle r_c^2\rangle_{0_1^+}$.
The excitation energies of both $0^+_2$ and $0^+_3$ exhibit a pronounced dip at
$N=90$, which can be attributed to the softness of the potential with respect to $\beta$
deformation in $^{150}$Nd. For lighter nuclei, i.e. toward spherical shapes, $0^+_2$ and
$0^+_3$ display the structure of two- and three-phonon states, respectively.  The axially
deformed $^{152,154,156}$Nd are characterized by strong prolate minima, stiffer
potentials, and the positions of  $\beta$ and $\beta \beta$ bands are
shifted to higher energies. This behavior of the $0^+_2$ state in Nd-Sm-Gd isotopes 
was also predicted in the phenomenological analysis of the transition between the 
vibrational, SU(5), and the rotational, SU(3), limits of the interacting boson model 
(IBM) \cite{Sch.78}. We note that, with the exception of the
very low $0^+_2$ state in $^{146}$Nd, the calculated excitation energies $E_{0^+_2}$
are also in quantitative agreement with experimental values \cite{NNDC}.

The microscopic calculation predicts a very interesting evolution of the isomer shift
$\langle r_c^2\rangle_{0_2^+}-\langle r_c^2\rangle_{0_1^+}$.
After a steep rise with neutron number for $N \leq 90$, the isomer
shift actually changes sign between $N=90$ and $N=92$, i.e. in $^{152,154,156}$Nd
the charge radius of the $0^+_2$ state is smaller than that of the ground state.
As function of the control parameter (number of neutrons), the isomer shift displays
a behavior characteristic for a first-order phase transition \cite{Iachello04}. 
Evidence for deformation crossing near the first-order shape-phase transition in 
$^{152-156}$Gd has recently been reported in a study that used experimental 
B(E2) values to extract the model-independent quadrupole shape invariants, which 
provide a measure of the $\beta$-deformation \cite{Wer.08}. Note that,
although the results correspond to a realistic calculation of ground-states and
collective excitation spectra of Nd nuclei, both isomer shifts, i.e.
$\langle r_c^2\rangle_{2_1^+}-\langle r_c^2\rangle_{0_1^+}$, and
$\langle r_c^2\rangle_{0_2^+}-\langle r_c^2\rangle_{0_1^+}$,
exhibit very sharp discontinuities at $N=90$. The first-order phase transition appears not to be
smoothed out by the finiteness of the nuclear system. In general, the characteristic behavior
of order parameters at the point of QPT is more pronounced than in the case
of Ising-type Hamiltonians
representing systems of interacting bosons, especially for a realistic number of bosons, i.e.
$N_B \approx 5-10$ for medium heavy nuclei. In the latter case the discontinuities are
smoothed out so that, qualitatively, a first-order phase transition might actually appear like a
second-order one \cite{Iachello04}.

Shape transitions and change in radii are also reflected in the transition
matrix elements of the electric monopole $\hat{T}(E0)$ operator \cite{Wood99}. In the study
of sharply rising $E0$ strength in transitional nuclei \cite{BWC.04}, based on a general
interacting-boson model Hamiltonian of Ising-type, it has been shown that
$0^+_2 \to 0^+_1$ transitions provide a clear signature of phase transitional
behavior in finite nuclei. In shape transition regions $\rho^2({E0; 0^+_2 \to 0^+_1})$
displays a steep rise, and then remains large for well deformed nuclei. Relative and
absolute $E0$ transition strengths on the transitional path between the X(5) solution and
the rigid rotor limit have recently been evaluated using the confined $\beta$-soft
(axially symmetric) rotor model (CBS) \cite{BKB.09}, and it has been shown that absolute 
$E0$ transition strengths are reduced with increasing potential stiffness toward zero in the 
rigid rotor limit.

\begin{figure}[htb]
\includegraphics[scale=0.225]{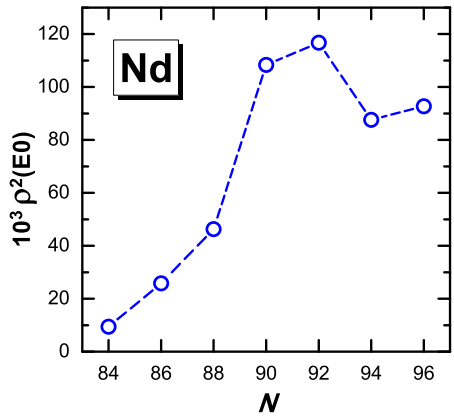}
\caption{\label{fig:rho-E0-Nd} (Color online) The calculated monopole transition
strength $\rho^2(E0;{0^+_2 \to 0^+_1})$ as a function of neutron number $N$ in Nd isotopes.}
\end{figure}

The E0 operator can be expressed in terms of single-nucleon degrees of freedom as:
$  \hat T({\rm E0})=\sum\limits_k e_kr_k^2$,
where $e_k$ is the charge of the $k$th nucleon, and $r_k$ is its position relative to the
center of mass of the nucleus. For the transition $0^+_2 \to 0^+_1$ the absolute $E0$ strength 
is defined
\begin{equation}
\rho^2(E0;{0^+_2 \to 0^+_1}) =\left|\frac{\langle0^+_2|\hat T({\rm E0})|0^+_1\rangle}
  {eR^2}\right|^2,
\end{equation}
where $R$ is the nuclear radius, $R\simeq1.2A^{1/3}$~fm. Fig.~\ref{fig:rho-E0-Nd} shows the
calculated values $\rho^2(E0;{0^+_2 \to 0^+_1})$ as a function of neutron number $N$. 
Bare charges have been used in the calculation, i.e. $e_p = e$ and $e_n = 0$.
The monopole transition strengths exhibits a markedly sharp increase toward the point of phase
transition at $N=90$, and the $\rho^2(E0;{0^+_2 \to 0^+_1})$ values remain rather large
in the well-deformed nuclei $^{152,154,156}$Nd, a result similar to that obtained in the
schematic interacting-boson model calculation of Ref.~\cite{BWC.04}. This behavior of
$\rho^2(E0;{0^+_2 \to 0^+_1})$ is characteristic for an order parameter at the point of
first-order QPT. In terms of absolute values, one notes that
even without introducing effective charges, the calculated $\rho^2(E0;{0^+_2 \to 0^+_1})$
are in qualitative agreement with the available experimental values for Sm and 
Gd nuclei \cite{Wood99}.

In conclusion, a microscopic calculation of observables related to 
order parameters for a first-order
nuclear QPT between spherical and axially deformed shapes have been performed.
Starting from self-consistent triaxial mean-field binding
energy maps in the $\beta - \gamma$ plane for a sequence of even-even Nd isotopes
with neutron number $N=84 - 96$, a set of observables has been computed
using collective wave functions obtained by diagonalization
of the corresponding  five-dimensional Hamiltonian for quadrupole vibrational and
rotational degrees of freedom. The energy gap between the ground 
state and the excited vibrational
states with zero angular momentum, the isomer shifts
$\langle r_c^2\rangle_{2_1^+}-\langle r_c^2\rangle_{0_1^+}$ and
$\langle r_c^2\rangle_{0_2^+}-\langle r_c^2\rangle_{0_1^+}$, and the monopole
transition strengths $\rho^2(E0;{0^+_2 \to 0^+_1})$ exhibit pronounced discontinuities
at $N=90$, characteristic of first-order QPT. Even though the calculation has been 
carried out for a finite number of nucleons, the phase transition does not appear
to be significantly smoothed out by the finiteness of the nuclear system. Together with
the results reported in Refs.~\cite{Ni.07,Li.09}, the present analysis has shown that the
microscopic framework based on universal energy density functionals
provides a fully consistent description of nuclear shape QPT in the rare-earth region 
around $N=90$. 
We thank R. F. Casten, F. Iachello and N. Pietralla for useful discussions.
This work was supported in part by MZOS - project 1191005-1010,
by the Major State 973 Program 2007CB815000, and
the NSFC under Grant No. 10775004.
The work of J.M, T.N., and D.V. was supported in part by
the Chinese-Croatian project "Nuclear structure far from stability". 
T. N. acknowledges support by the Croatian National Foundation for 
Science, Higher Education and Technological Development.



\end{document}